\documentclass[12pt,preprint]{aastex}
\begin{document}
\newcommand{\calu}{{\cal U}}
\newcommand{\calq}{{\cal Q}}
\newcommand{\bx}{{\rm \bf x}}
\newcommand{\bk}{{\bar{\kappa}}}
\title{The Fate of Non-Radiative Magnetized Accretion Flows:
Magnetically-Frustrated Convection} 
\author{Ue-Li Pen}
\affil{Canadian Institute for Theoretical Astrophysics, University of
Toronto, M5S 3H8, Canada; pen@cita.utoronto.ca}
\author{Christopher D. Matzner}
\affil{Department of Astronomy and Astrophysics, University of
Toronto, M5S 3H8, Canada;  
matzner@cita.utoronto.ca} 
\author{Shingkwong Wong}
\affil{Department of Physics, National Taiwan University;
wshingkw@cita.utoronto.ca}
\begin{abstract}
We present a scenario for non-radiative accretion onto the
supermassive black hole at the galactic center.  Conducting MHD
simulations with $1400^3$ grid zones that break the axial and
reflection symmetries of earlier investigations and extend inward from
the Bondi radius, we find a quasi-hydrostatic radial density profile
$\rho \propto r^{-0.72}$ with superadiabatic gradient corresponding to
an $n \sim 0.72$ polytrope.  Buoyancy generated by magnetic
dissipation is resisted by the same fields so effectively that energy
is advected inward: a state of magnetically-frustrated convection.
This scenario is consistent with observational constraints on
energetics andouter boundary conditions.
\end{abstract}

\section{Introduction}\label{intro}

Stellar dynamical measurements indicate a black hole with mass $M_{\rm
BH}=2.4\times 10^6 M_\odot$ \citep{1997MNRAS.291..219G,
1998ApJ...509..678G} at the location of Sgr A* in the galactic center.
Apart from its origin, a puzzling aspect of this object -- and others
like it in nearby galaxies -- is its low X-ray luminosity, given the
gaseous environment.  High resolution X-ray imaging
\citep{2001astro.ph.02151P} shows hot gas with temperatures $T= 2$ keV
at densities near $n_e=130 {\rm cm}^{-3}$ within 1'' of Sgr A*.  In
\cite{1952MNRAS.112..195B}'s theory, gas within the gravitational
region of influence (Bondi radius)
\begin{equation}\label{rB}
r_B\equiv \frac{GM_{\rm BH}}{c_s^2} \simeq 0.03 {\rm pc}
\end{equation}
falls inward, approaching free-fall.  This scale is barely resolved by
{\em Chandra} at the galactic center.  The natural mass accretion rate
is of order $4\pi\lambda r_B^2 \rho c_s$ if the background density is
$\rho$; $\lambda = 0.25$ for a monatomic gas dominated by thermal (as
opposed to magnetic) pressure.  If matter is converted into radiation
at an efficiency $\eta$ by the black hole, the luminosity $L\simeq
2\times 10^{40} (\eta/10\%)$ erg/s.  Reported instead
\citep{2001astro.ph.02151P} is a source, potentially the hole, with a
luminosity of $2.4\times 10^{33}$ erg/s -- $10^7$ times fainter than
this estimate.  Even the advection of thermal energy across $r_B$ at
the Bondi rate would incur $\sim 2\times 10^{36}$ erg/s, exceeding the
observations by three orders of magnitude.  Because Bondi assumed
spherical adiabatic flow of an ideal gas, many factors may be at work
in this immense discrepancy.

First, the flow could be incredibly sporadic and we may have caught it
in an off moment.  However, the dynamical time $r_B/c_s$ is only $\sim
50$ years, comparable to the history of radio observations and only a
few times longer than the X-ray observations.

Second, it had been suggested \citep[as in the ADAF model
of][]{1994ApJ...428L..13N} that fluid can accrete at the Bondi rate
without radiating ($\eta \ll 10\%$), which might be possible if
electrons coupled to protons only by Coloumb collisions.  However,
observations of linear submillimeter polarization
\citep{2000GCNew..11....4B} are interpreted
\citep{2000ApJ...545L.117M} to imply an accretion rate far below
Bondi's prediction.  Moreover, Bondi-rate influx of electron thermal energy
alone would exceed limits by $\sim 10^{2.7}$.

Third, Bondi accretion passes through a sonic point only if the
effective adiabatic index is smaller than $5/3$.  Gas with
$\gamma_{\rm eff}=5/3$ accretes subsonically at all radii, and if
$\gamma_{\rm eff}>5/3$ then quasi-hydrostatic settling flow is
expected, becoming sonic just above the Schwarzschild radius $r_{\rm Sch}$
\citep[][]{1978A&A....70..583B}.  Monatomic gas with $\gamma = 5/3$
has $\gamma_{\rm eff}>5/3$ if viscous or magnetic dissipation
generates heat (which is not radiated) during inflow
\citep[][]{1978ApJ...226.1041C,1981ApJ...246L..15S}.  Therefore, rapid
inflow of the Bondi or ADAF type is unstable to motions that reverse
the inflow of some fluid elements -- potentially reducing
$\dot{M}_{\rm BH}$ far below the Bondi estimate.  The ADIOS model of
\cite{1999MNRAS.303L...1B} posits axial outflows cancelling the
equatorial inflow.  Similarly, the CDAF model of
\cite{2000ApJ...539..809Q} invokes a rotating convecting atmosphere
surrounding the hole.  \citeauthor{2002ApJ...566..137I} (2002,
hereafter IN) and \cite{2001astro.ph..4113G} discuss a similar,
nonrotating, convecting flow (\S \ref{phyisicalinterp}). 

Fourth and fifth, rotation and magnetic fields can both strongly
affect the flow and interact with one another and with other physical
effects in non-trivial ways. 

If the angular frequency is $\Omega_B$ at the Bondi radius,
conservation of angular momentum would bring it into orbit at a Kepler
radius $r_K\simeq \Omega_B^2 r_B^3/c_s^2$.  If cooling permits the
formation of a thin disk, this signals the onset of efficient
radiation ($\eta\simeq 10\%$ as used above).  If angular momentum
transport is weak, however, this can be a limiting step in accretion.
\cite{2003astro.ph..2420N} has recently suggested a cold disk onto
which the hot flow condenses, accreting sporadically with a long duty
cycle.  The disk of stars inferred by \cite{2003astro.ph..3436L} could
have arisen this way.  Recall, however, condensation onto this
disk would exceed constraints if it emitted X-rays.

Similarly, magnetic fields can grow in strength as they are dragged
inward by the flow.  Accretion will shear embedded fields to become
radial.  An initial field, whether uniform or tangled
\citep{1973ApJ...185...69S}, will be stretched toward the
split-monopole (hedgehog) configuration in which $B^2 \propto r^{-4}$
(e.g., IN), incurring a centrally divergent energy.  Although radial
fields exert no net force, this configuration is unstable and
unattainable \citep[][]{1971reas.book.....Z}.  Instead, inflow can
stall at a magnetic turnaround radius $r_{\rm mag} \equiv G
M/v_A(r_{\rm mag})^2$, where $v_A(r)$ is the Alfv\'en velocity.  Mass
inflow will be limited by the rate of magnetic energy dissipation via
reconnection, by interchange instabilities, or by inflow along open
field lines.

This contrasts with the widely-held picture of magnetic fields
enhancing accretion through angular momentum transport, either via
turbulent stresses \citep{1991ApJ...376..214B} or via
magnetocentrifugal winds \citep{1982MNRAS.199..883B}.  Also, if fields
are generated locally through convection or magnetorotational
instability, then they are limited in strength to partial
equipartition with kinetic energy.  Fields strengthened by inflow can
reach equipartition with the gravitational potential, which can be a
larger value (\S \ref{results}).

We pause here to note the critical importance of the density profile
in the phenomenology and viability of models; see also
\cite{2003astro.ph..4099Q} and \cite{2001astro.ph..4113G}.  An
atmosphere with $\rho \propto r^{-n}$ and, generically, $T\propto
r^{-1}$, has a bolometric free-free luminosity $L\propto r^3 \rho^2
T^{1/2} \propto r^{(5-4n)/2}$ (or for a single frequency, $L_\nu
\propto r^3\rho^2 T^{-1/2}\propto r^{7/2-2n}$), so long as $T\gtrsim
3$ keV.  Thus profiles with $n>5/4$ (or $n>7/4$) suffer a singularity in
the bolometric (or single-frequency) luminosity.  If $n<5/4$, emission is
dominated at outer radii; this is consistent with the small contrast
in {\em Chandra} images.

Similarly, the thermal time varies as $T^{1/2}/\rho\propto r^{n-1/2}$
whereas the inflow time $r/v = 4\pi r^3 \rho/\dot{M}_{\rm BH}\propto
r^{3-n}$.  Thus, flows with $n<7/4$ are dynamically nonradiative.
Finally, the local Bondi accretion rate varies as $r^2\rho c_s\propto
r^{3/2-n}$: flows with $n<3/2$ accrete at $\dot{M} \simeq \dot{M}_{B}
(R_B/R_{\rm Sch})^{3/2-n}$.

Hydrostatic models are characterized by a polytropic index $n =
1/(\gamma_{\rm eff}-1)$, which describes the correlation between
density and pressure $p\propto \rho^{\gamma_{\rm eff}} =
\rho^{1+1/n}$.  In a Keplerian potential, $\rho\propto r^{-n}$ as used
above.  When $\gamma_{\rm eff} = 5/3$, both hydrostatic and Bondi-like
(e.g., ADAF) flows have $n=3/2$, shallow enough to be advective but
too steep to avoid the luminosity constraint.  To force a shallower
density profile requires additional pressure, either through rotation,
magnetic fields, or an entropy inversion.  The latter is convectively
unstable.

CDAFs pass this test by achieving $n=1/2$ in a combination of
rotational, turbulent, and thermal support.  However, the
\cite{2000ApJ...539..809Q} model requires the inward angular momentum
transport due to buoyant convection to exceed the outward transport
due to magnetic fields.  This may be possible for fields due solely to
magnetorotational instability \citep{2002ApJ...577..295N}, but it is
unlikely a property of field strengthened by shear in inflow.

More seriously, any model invoking rotational support is plagued by
outer boundary conditions.  Outside $r_B$ one expects solid-body
rotation on radial shells, implying low specific angular momentum
($j$) near the axis.  Sufficiently low-$j$ gas can fall directly
within the innermost stable orbit; this comprises a fraction $\sim
(r_S/r_K)^{1/2}$ of the total.  If $j$ is somewhat higher, fluid falls
inward to its own Kepler radius and shocks; the resulting pressure
gradients drive a quadrupolar outflow of low-$j$ material along the
equator.  This has been observed in nonmagnetic simulations by
\citet{2003ApJ...582...69P} and by us.  It could be avoided if the
inner boundary supplied a strong jet that interfered with axial
accretion, as occurs in protostars \citep{2000ApJ...545..364M}, but
suggestions of such behavior in {\em Chandra} images are recent and
tentative \citep{2002AAS...201.3105M}.
Similarly, models that invoke rotational support to stem accretion
require that rotational support remains important all the way out to
$r_B$ -- implying an asymmetry to the X-ray images which is not
observed.

How can gas establish the flattened density profile required for
low-luminosity accretion, while being fed
low-angular-momentum, lightly magnetized ($\beta>10$ for $B<1$ mG)
material at $r_B$?  IN and \cite{2001astro.ph..4113G} offer
a solution to this conundrum (\S \ref{phyisicalinterp}); our
simulations suggest a different answer. 

\section{Simulations} \label{thiswork} 

In order to further explore the interaction between infall, rotation,
magnetic fields, and buoyancy, we conduct MHD simulations with
1400$^3$ zones arrayed in a uniform Cartesian grid, the largest MHD
simulations to date. These were performed on the CITA McKenzie
cluster: 512 Pentium-4 Xeon processors running at 2.4
GHz \citep{2003astro.ph..5109D}. At this resolution, each full
dimensional sweep corresponding to two timesteps took 40 seconds.  The
code \citep{2003astro.ph..5088P} is based on a 2nd order accurate (in
space and time) high resolution Total-Variation-Diminishing (TVD)
algorithm. It explicitly conserves the sum of kinetic, thermal and
magnetic energy; hence magnetic dissipation (at the grid scale) heats
gas directly.  No explicit resistivity or viscosity is added, and
reconnection and shocks occur through the solution of the flux
conservation laws and the TVD constraints.  Magnetic flux is conserved
to machine precision by storing fluxes perpendicular to each cell
face.

Inner boundary conditions are imposed on a cube of width 24 grid
cells.  Interior to the largest inscribed sphere within this cube,
gravitational forces are turned off.  At the end of each time step
magnetic fields in the cubical region are relaxed to the vacuum solution,
permitting rapid reconnection in the interior zone.  The Alfv\'en
speed is matched to the circular speed at the surface of this region
by the removal of matter, and the sound speed is matched to the same
value by the adjustment of temperature.  The pressure of this matter
is always smaller than that of infalling material, and we never
observe spurious outflows.

Whereas many prior simulations have started with a disklike, rotation
dominated geometry and no low-$j$ material, we wished to investigate
the potential role of axial infall.  We attempt to separate generic
physical effects from artefacts of  boundary conditions.  This is
challenging, since the energy available at the center of the
simulation is always larger than in any other region.  Results
may depend sensitively on the choice of inner boundary, and it is
not feasible to simulate directly from $r_B$ to
$r_{\rm Sch}$.  We search instead for scaling relations connecting
these radii, bearing in mind the possibility of strong
reactions such as outflows from an unresolved inner region.

For this reason, and since we are interested primarily in the effect
of external parameters, most of our runs have worked inward from the
Bondi radius.  Simulations far inside $r_B$ elucidate local phenomena,
but cannot easily be matched to their exterior environment.  At each
step, the outer 20 grid cells are replaced with the values from the
initial conditions to enforce the continued inflow of new material.  We
avoid threading the outer boundary with any magnetic flux by adding
flux lines to the fluid inside a region 1/2--3/4 of the
box size, with an admixture of random field loops and large loops that
thread the whole box.  (Most of the energy is in the coherent loops.)
We maintain a mean value $\beta=10$ in the flux generation region.  To
estimate $r_{\rm mag}$, we assume $B\propto r^{-2}$ as in the
split-monopole.  This gives $r_{\rm mag}\simeq 100$ grid cells.  The
estimate is not rigorous: fields are added in a cubical region;
rotation changes the shearing rate, and $B\propto r^{-2}$ is not
perfectly achieved.

To avoid immediate energetic feedback from the central regions due to
the initial conditions, we start with an empty interior (an evacuated
sphere of half the box radius), and let matter fall in.  We then watch
the evolution away from Bondi to draw conclusions about the
fate of the flow.  In grid units, box length is 1400, and $GM=700$.
We set $c_s(r=\infty)=1$, and $\rho(r=\infty)=1$.  The Bondi radius
was set at $r_B=700$, touching the closest box edge.  The simulations
ran successfully for 6000 time steps to $t=650$, or 1.5 free fall
times from $r_B$.  At that point, magnetic
fields squeezed out along the midplane to the outer boundary, leading
to numerical instabilities.

With a resolution of $10^{2.85}$ radial zones, of which $10^{1.51}$ is
used for central and outer boundary conditions, we have 1.34 decades
of scale in which to arrange the Kepler radius $r_K$ and the magnetic
turnaround radius $r_{\rm mag}$.  We have performed runs that are
initially rotationally supported ($r_{\rm mag}<r_K$) and purely
rotationally supported ones ($r_{\rm mag} = 0$), as well as magnetized
but nonrotating initial conditions ($r_K = 0$).  The fluid is
initialized with solid body rotation on shells, and constant specific
angular momentum on cones of constant polar angle.  The production
simulation (fig. \ref{fig:mag}) chose $r_K=336$, $r_{\rm mag} = 100$.
To further break any discrete symmetries, the velocity field was
modulated at the 5\% level at multipoles up to $l=2$.

These simulations differ from previous ones in two important
ways.  First, magnetic and viscous dissipation occur only at the grid
scale (or inner boundary); no enhanced diffusivity or viscosity is
applied.  Second, we break the alignment usually assumed between
magnetic and rotational symmetry axes.  This was accomplished by
introducing large-scale flux loops, misaligned with respect to the
rotational axis, that are dragged toward the central object.

\subsection{Results}\label{results}
\begin{figure}
\epsscale{0.9}
\plotone{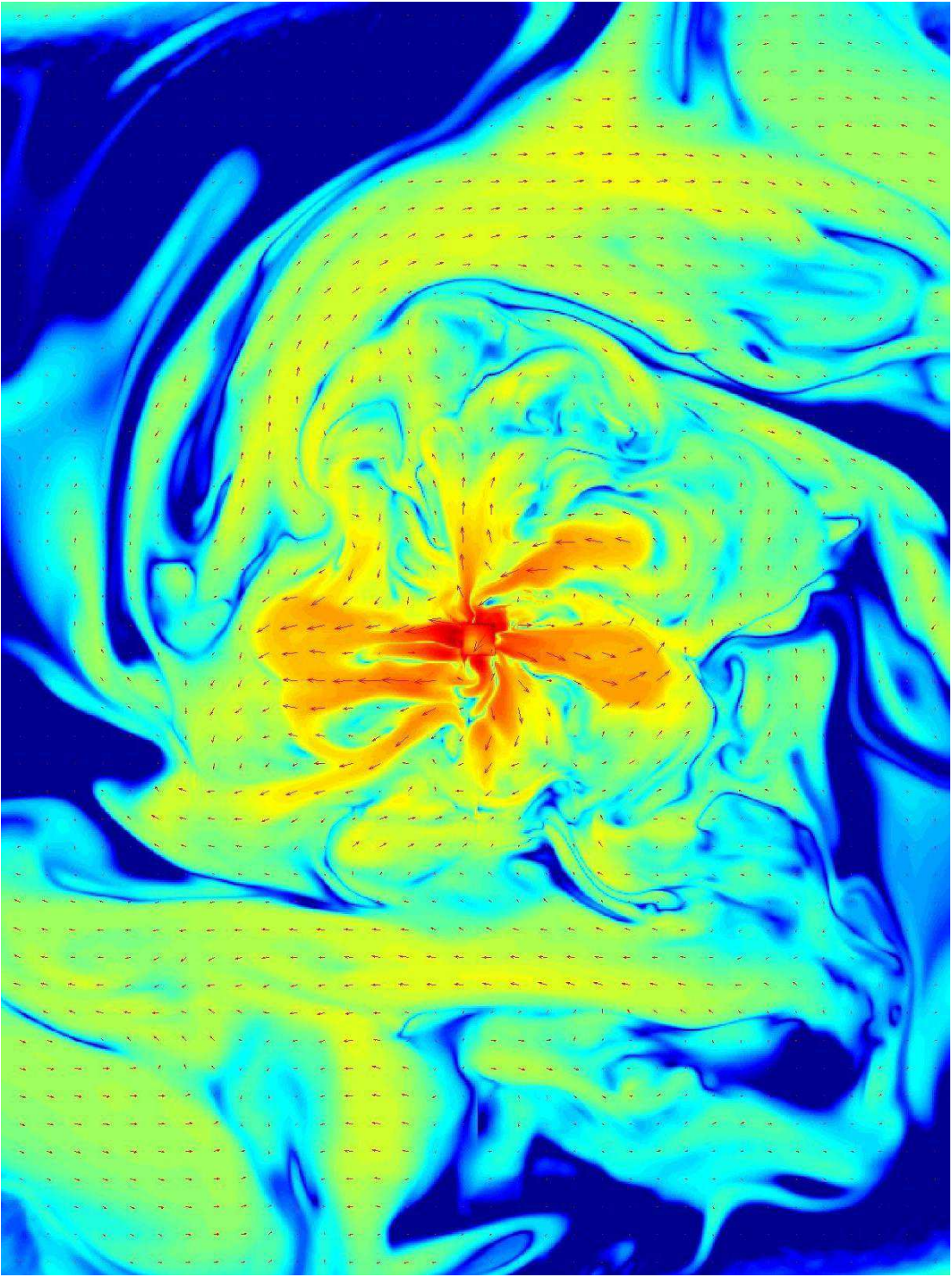}
\caption{Magnetic field structure: midplane ($x-y$) slice. Magnetic
  pressure is shown along with projected magnetic field vectors. The
  inner half of this picture corresponds to figure \ref{fig:stresses}.}
\label{fig:mag}
\end{figure}

\begin{figure}
\epsscale{1}
\plotone{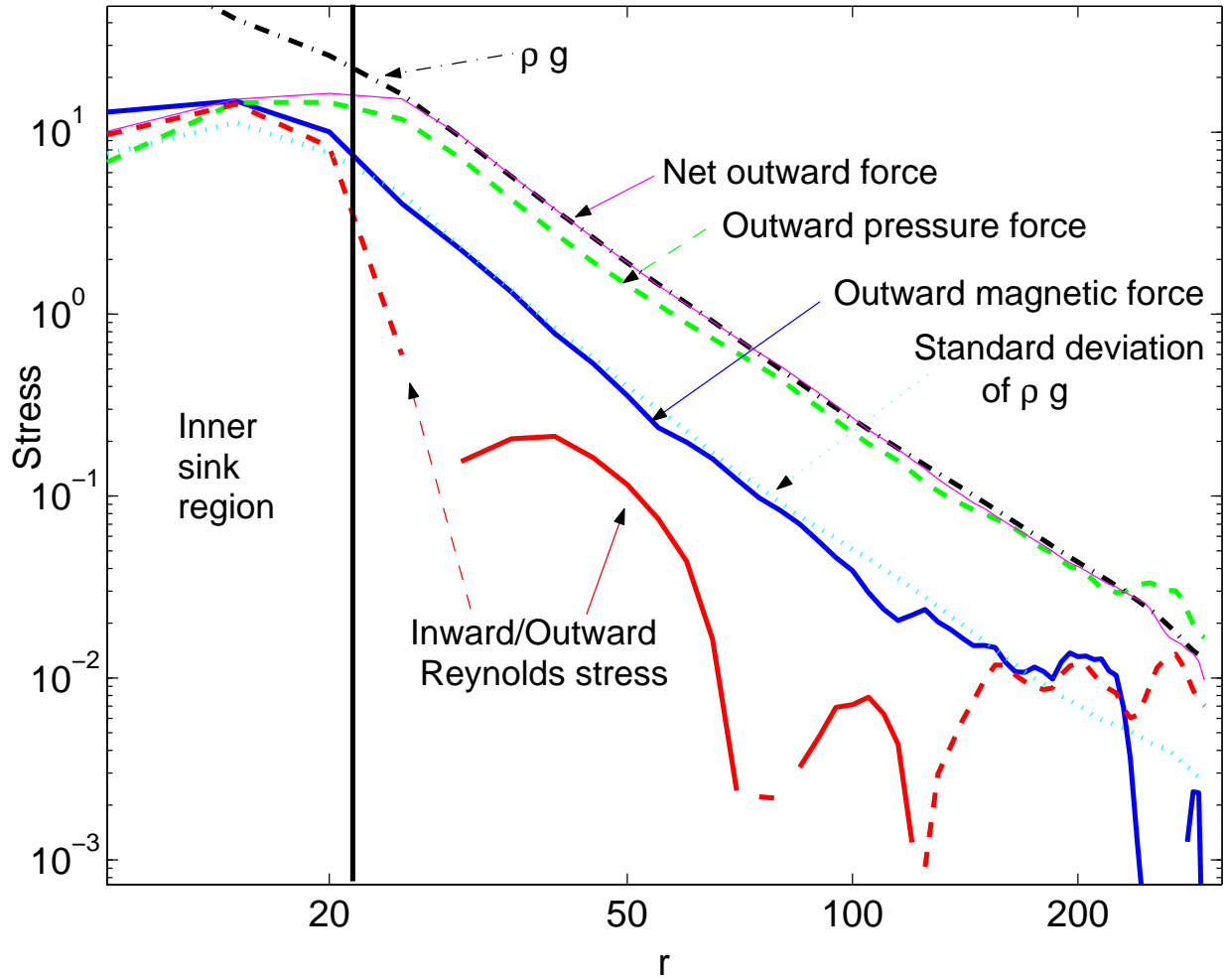}
\caption{Run of supporting stresses (radial components of momentum
fluxes), averaged on radial shells and compared to the local
gravitational force per unit volume. Also plotted is the standard
deviation of gravitational force per unit volume on radial shells. }
\label{fig:stresses}
\end{figure}

Simulations were run about one dynamical time at $r_B$.  Although this
does not give us any handle on the long-term evolution of the flow
once its inner behavior begins to alter the conditions outside $r_B$,
it does yield a picture of how the inner flow responds -- for many
inner dynamical times -- to initial and boundary conditions.  A
snapshot of the resulting magnetic field structure is shown in
\ref{fig:mag}, and various stresses balancing gravity are plotted
versus radius in figure \ref{fig:stresses}.

\clearpage

\clearpage

At this point in the simulation, a central hydrostatic region
supported by gas pressure -- not rotation or magnetic fields -- has
grown outward.  Within the region plotted in figure
\ref{fig:stresses}, $\rho \propto r^{-n}$ for $n\simeq 0.72$ also,
$P\propto r^{-1.51}$ (so that $T\propto r^{-0.79}$, whereas $r^{-1}$
was expected), and magnetic pressure $P_{\rm mag} \sim 10^{-1.5} P$.
Gas pressure gradients dominate magnetic stresses by a factor $\sim
10$; Reynolds stresses, including rotation and inflow, are smaller
still (and switch direction).  Rotation is roughly one tenth the
Kepler rate, despite that the entire region is a factor of two inside
the initial $r_K$.  The ratio of Alfv\'en to inflow velocities is
similarly $\sim 10$, indicative of magnetic braking.

In contrast to Bondi and ADAF-type flows, inflow is very subsonic;
correspondingly, the mass accretion rate is strongly suppressed
relative to Bondi's estimate.  However, the accretion rate does agree
with the Bondi rate derived from conditions at the inner boundary.
This state is not rotationally-supported like the CDAF.  It resembles
in some ways the ``CDBF'' of IN and \cite{2001astro.ph..4113G}; but see \S
\ref{phyisicalinterp} for differences.

To test the role of magnetic fields in the clogged inflow, we suddenly
turned the magnetic fields off, and evolved the fluid for a dynamical
time.  The flow returned to Bondi's solution. 

\section{Physical Interpretation}\label{phyisicalinterp}
The fluid can be modelled as one in quasi-hydrostatic equilibrium with
a polytropic index $\gamma_{\rm eff}\simeq 2.25$.  Since the adiabatic
index is $\gamma = 5/3$, this represents a strongly superadiabatic
state.  In the usual description of entropy-driven convection, this
can only occur when convective velocities approach the sound speed,
which, in a power-law atmosphere, is roughly free-fall.

Saturation at a constant Mach number is a feature of
\cite{2000ApJ...539..809Q}'s CDAF model and the CDBF model of IN, both
of which have $n=1/2$ and fall within class II of
\cite{2001astro.ph..4113G}'s classification of self-similar
nonradiative flows.  This slope is clearly flat enough to satisfy all
of the observational constraints discussed in \S \ref{intro}.  The
value of $n=1/2$ derives from assuming a positive convective
luminsoity $L_{\rm conv}$ -- i.e., that the gravitational energy
released bubbles out through the flow rather than being dragged inward
(as it is in Bondi and ADAF flows).  If $c_s$ scales with the Kepler
speed $v_K$, then $L_{\rm conv} \propto r^2 \rho v_K^3 {\cal M}
\propto r^{1/2-n} {\cal M}$, where ${\cal M}$ is the turbulent Mach
number.  Constancy of $L_{\rm conv}$, required by steady state,
implies ${\cal M}\propto r^{n-1/2}$.  If $n>1/2$, ${\cal M}$ rises
outward until it saturates at $\sim 1$; thereafter,
$n=1/2$. 

Our simulation shows $n=0.72$ and mildly subsonic convective motions:
${\cal M}\simeq 0.4$ -- not increasing outward at $r^{0.22}$ as implied
by the above argument.  This has prompted us to calculate the total
flow ($L_{\rm conv}$) of magnetic, thermal, and gravitational
energy.  We find it to be {\em inward} and small: $L_{\rm conv}\simeq
-0.04 \times (4\pi r^2 \rho v_K^3)$.  

How can subsonic convection persist in the presence of strong
superadiabatic gradients?  And, how can the net flow of energy be
inward despite the existence of convective motions?  Since the
detailed treatment of magnetic field structures is the only physical
effect distinguishing our final state from ordinary gaseous
convection, the simulation results must represent a state of {\em
magnetically-frustrated convection} in which magnetic shear stresses
oppose buoyant motions.  This is verified by the equipartition between
magnetic stresses and buoyant stresses (variations of $\rho g$ at
fixed $r$) seen in figure \ref{fig:stresses}.  Correspondingly,
kinetic energy is suppressed by a factor of three compared to the free
energy available in density fluctuations (variations of $\rho GM/r$ at
fixed $r$).  For further corroboration we note an anticorrelation
between magnetic and buoyant stresses (cross-correlation coefficient
-0.2) and also the relaxation to Bondi flow when fields were removed.
An analogy to magnetic frustration is low-Reynolds-number convection,
in which viscosity slows or halts buoyancy.  (That said, energy inflow
may be a feature of our inner boundary conditions -- a question for
further study.)

The coincidence in slope between gas and magnetic pressures most likely
arises from the growth of magnetic fields to balance buoyancy: since
buoyancy scales with gas pressure in the presence of a strong
superadiabatic gradient, so must the magnetic stress. 

One can discuss the plausible scenarios of the global gasdynamics of
the Sgr A* region.  In addition to the ambient gas, stellar winds may
inject a larger mass flux into the region than would be lost by even
ideal Bondi accretion \citep{1999ApJ...517L.101Q}.  If the natural
accretion rate is sufficiently slow, the injected mass could
conceivably push the ambient gas outwards.  Similarly, the cooling
time at the Bondi radius is approximately $10^5$ years, and on this
time scale matter has to either flow in, or be pushed out.

\section{Conclusions}

We have presented a new picture of the hot plasma in the vicinity of
the galactic center black hole based on the results of new very large
three dimensional MHD simulations.  We find that the code produces
results consistent with observational constraints, including boundary
conditions and total luminosity.  Fluid remains in quasi-static
equilibrium supported primarily by thermal pressure, with a radial
density profile $\rho\propto r^{-0.72}$ in reasonably spherical
symmetry.  This implies a significant entropy inversion; we believe 
buoyancy is impeded by subdominant magnetic fields. 

A distinguishing feature of the flow is that energy is advected {\em
inward}, despite convection, albeit quite slowly compared to the local
estimate $4\pi r^2 \rho(r) v_K(r)^3$.  The same is true of central
regions of Bondi flow, where $\dot{E}= -\dot{M} c_s(r>t_B)^2/
(\gamma-1)$.  Unlike Bondi, the $\dot{M}$ is also much slower than the
dynamical accretion rate.

This feature provides a point of contrast with related suggestions for
the Sgr A* flow: CDAF \citep{2000ApJ...539..809Q} and CDBF (IN), both
of which obey $\rho\propto r^{-1/2}$ and contain different levels of
rotational support.  The $1/2$-law in these models is a consequence of
a positive outward convective luminosity, plus the saturation of
convective motions at constant Mach number \citep[see
\S~\ref{phyisicalinterp} and][]{2001astro.ph..4113G}.  However, the
density profile is essentially unconstrained when $\dot{E}$ and
$\dot{M}$ (as well as the angular momentum flux $\dot{J}$) are all
effectively zero \citep[][his class-IV flow]{2001astro.ph..4113G}.

The influx of energy also provides a method of distinguishing
magnetically-frustrated flows from the other models even if the
central density profile cannot be probed.  In CDAFs and CDBFs, energy
must be convected outward and will heat gas outside the Bondi radius
at a rate comparable to $\dot{M} c_{s}^2\sim 10^3 L_\odot$.  If
present, effects of this heating on the hot gas surrounding the hole
may be visible.  On the other hand, winds from massive stars in the
same volume may overwhelm inward advection.

\cite{1999MNRAS.303L...1B} criticize ADAFs on the basis that they
possess a positive Bernoulli function (${\cal B} >0$), and that this
predisposes them to outflow (realized in their ADIOS model).  Note
however that $\cal B$ is positive (though small) in Bondi flow, where
it is set by the external pressure; however, no outflow develops.  We
find ${\cal B}>0$ in the magnetically-frustrated flow as well.  We do
observe upwelling, but it takes the form of an overall quadrupolar
circulation with inflow along the axis and outflow along the equator.
There is no indication that this upwelling becomes supersonic at any
radius.

We speculate that non-radiative MHD flows are in general very inefficient
at accreting.  The corollary is that accretion only occurs when the cooling
time is short.

\acknowledgments   

This research was funded by NSERC, by the Canadian Foundation for
Innovation, and by the Canada Research Chairs program (for CDM).  We
thank Phil Arras, David Ballantyne, Robert Fisher, Chris McKee, and
an anonymous referee for comments.

\bibliography{penbib}

\begin{thebibliography}{28}
\expandafter\ifx\csname natexlab\endcsname\relax\def\natexlab#1{#1}\fi

\bibitem[{{Baganoff} {et~al.}(2001){Baganoff}, {Maeda}, \&
  {Morris}}]{2001astro.ph.02151P}
{Baganoff}, F., {Maeda}, Y., \& {Morris}, M. 2001, in eprint
  arXiv:astro-ph/0102151, 02151--+

\bibitem[{{Balbus} \& {Hawley}(1991)}]{1991ApJ...376..214B}
{Balbus}, S.~A. \& {Hawley}, J.~F. 1991, \apj, 376, 214

\bibitem[{{Begelman}(1978)}]{1978A&A....70..583B}
{Begelman}, M.~C. 1978, \aap, 70, 583

\bibitem[{{Blandford} \& {Begelman}(1999)}]{1999MNRAS.303L...1B}
{Blandford}, R.~D. \& {Begelman}, M.~C. 1999, \mnras, 303, L1

\bibitem[{{Blandford} \& {Payne}(1982)}]{1982MNRAS.199..883B}
{Blandford}, R.~D. \& {Payne}, D.~G. 1982, \mnras, 199, 883

\bibitem[{{Bondi}(1952)}]{1952MNRAS.112..195B}
{Bondi}, H. 1952, \mnras, 112, 195

\bibitem[{{Bower}(2000)}]{2000GCNew..11....4B}
{Bower}, G. 2000, Galactic Center Newsletter, 11, 4

\bibitem[{{Cowie} {et~al.}(1978){Cowie}, {Ostriker}, \&
  {Stark}}]{1978ApJ...226.1041C}
{Cowie}, L.~L., {Ostriker}, J.~P., \& {Stark}, A.~A. 1978, \apj, 226, 1041

\bibitem[{{Dubinski} {et~al.}(2003){Dubinski}, {Humble}, {Pen}, {Loken}, \&
  {Martin}}]{2003astro.ph..5109D}
{Dubinski}, J., {Humble}, R., {Pen}, U., {Loken}, C., \& {Martin}, P. 2003,
  ArXiv Astrophysics e-prints

\bibitem[{{Genzel} {et~al.}(1997){Genzel}, {Eckart}, {Ott}, \&
  {Eisenhauer}}]{1997MNRAS.291..219G}
{Genzel}, R., {Eckart}, A., {Ott}, T., \& {Eisenhauer}, F. 1997, \mnras, 291,
  219

\bibitem[{{Ghez} {et~al.}(1998){Ghez}, {Klein}, {Morris}, \&
  {Becklin}}]{1998ApJ...509..678G}
{Ghez}, A.~M., {Klein}, B.~L., {Morris}, M., \& {Becklin}, E.~E. 1998, \apj,
  509, 678

\bibitem[{{Gruzinov}(2001)}]{2001astro.ph..4113G}
{Gruzinov}, A. 2001, ArXiv Astrophysics e-prints, 4113

\bibitem[{{Igumenshchev} \& {Narayan}(2002)}]{2002ApJ...566..137I}
{Igumenshchev}, I.~V. \& {Narayan}, R. 2002, \apj, 566, 137

\bibitem[{{Levin} \& {Beloborodov}(2003)}]{2003astro.ph..3436L}
{Levin}, Y. \& {Beloborodov}, A.~M. 2003, ArXiv Astrophysics e-prints, 3436

\bibitem[{{Matzner} \& {McKee}(2000)}]{2000ApJ...545..364M}
{Matzner}, C.~D. \& {McKee}, C.~F. 2000, \apj, 545, 364

\bibitem[{{Melia} {et~al.}(2000){Melia}, {Liu}, \&
  {Coker}}]{2000ApJ...545L.117M}
{Melia}, F., {Liu}, S., \& {Coker}, R. 2000, \apjl, 545, L117

\bibitem[{{Morris} {et~al.}(2002){Morris}, {Baganoff}, {Howard}, {Maeda},
  {Bautz}, {Feigelson}, {Brandt}, {Chartas}, {Garmire}, \&
  {Townsley}}]{2002AAS...201.3105M}
{Morris}, M.~R., {Baganoff}, F.~K., {Howard}, C.~D., {Maeda}, Y., {Bautz}, E.,
  {Feigelson}, M., {Brandt}, N., {Chartas}, G., {Garmire}, G., \& {Townsley},
  L. 2002, American Astronomical Society Meeting, 201, 0

\bibitem[{{Narayan} {et~al.}(2002){Narayan}, {Quataert}, {Igumenshchev}, \&
  {Abramowicz}}]{2002ApJ...577..295N}
{Narayan}, R., {Quataert}, E., {Igumenshchev}, I.~V., \& {Abramowicz}, M.~A.
  2002, \apj, 577, 295

\bibitem[{{Narayan} \& {Yi}(1994)}]{1994ApJ...428L..13N}
{Narayan}, R. \& {Yi}, I. 1994, \apjl, 428, L13

\bibitem[{{Nayakshin}(2003)}]{2003astro.ph..2420N}
{Nayakshin}, S. 2003, ArXiv Astrophysics e-prints, 2420

\bibitem[{{Pen} {et~al.}(2003){Pen}, {Arras}, \& {Wong}}]{2003astro.ph..5088P}
{Pen}, U., {Arras}, P., \& {Wong}, S. 2003, ArXiv Astrophysics e-prints

\bibitem[{{Proga} \& {Begelman}(2003)}]{2003ApJ...582...69P}
{Proga}, D. \& {Begelman}, M.~C. 2003, \apj, 582, 69

\bibitem[{{Quataert}(2003)}]{2003astro.ph..4099Q}
{Quataert}, E. 2003, ArXiv Astrophysics e-prints, 4099

\bibitem[{{Quataert} \& {Gruzinov}(2000)}]{2000ApJ...539..809Q}
{Quataert}, E. \& {Gruzinov}, A. 2000, \apj, 539, 809

\bibitem[{{Quataert} {et~al.}(1999){Quataert}, {Narayan}, \&
  {Reid}}]{1999ApJ...517L.101Q}
{Quataert}, E., {Narayan}, R., \& {Reid}, M.~J. 1999, \apjl, 517, L101

\bibitem[{{Scharlemann}(1981)}]{1981ApJ...246L..15S}
{Scharlemann}, E.~T. 1981, \apjl, 246, L15

\bibitem[{{Shapiro}(1973)}]{1973ApJ...185...69S}
{Shapiro}, S.~L. 1973, \apj, 185, 69

\bibitem[{{Zeldovich} \& {Novikov}(1971)}]{1971reas.book.....Z}
{Zeldovich}, Y.~B. \& {Novikov}, I.~D. 1971, {Relativistic astrophysics. Vol.1:
  Stars and relativity} (Chicago: University of Chicago Press, 1971)

\end{thebibliography}
\bibliographystyle{apj}

\appendix

\end{document}